\documentstyle[aps,prl]{revtex}
\input{epsf}
\draft
\begin{document}
% for two column  activate the line below...
\twocolumn[\hsize\textwidth\columnwidth\hsize\csname@twocolumnfalse\endcsname
%LA-UR-98-4085
\title
{Structure of vortex liquid phase in irradiated
Bi$_2$Sr$_2$CaCu$_2$O$_{8-\delta}$ crystals}
\author{N.~Morozov$^a$, M.P.~Maley$^a$, L.N.~Bulaevskii$^a$,
V.~Thorsm$\o$lle$^a$\\ A.E.
Koshelev$^b$, A.~Petrean$^b$ and W.K.~Kwok$^b$}
\address{$^a$Superconductivity Technology Center,
Los Alamos National Laboratory, Los Alamos, NM 87545 \\
$^b$ Materials Science Division, Argonne National Laboratory, Argonne,
Illinois 60439}
\date{\today}
\maketitle
\begin{abstract}
The $c$-axis resistivity in irradiated and in pristine
Bi$_2$Sr$_2$CaCu$_2$O$_{8-\delta}$ crystals is measured as a function
of the in-plane magnetic field component at fixed out-of-plane
component $B_{\bot}$ in the vortex liquid phase at $T=67$ K. From this
data we extract the dependence of the phase difference correlation
length inside layers on $B_{\bot}$ and estimate the average length of
the vortex line segments confined inside columnar defects as a
function of the filling factor $f=B_{\bot}/B_{\Phi}$.  The maximum
length, about 15 interlayer distances, is reached near $f\approx
0.35$.
\end{abstract}
% for two column  activate the line below...
]

The structure of the vortex liquid in highly anisotropic layered
superconductors with columnar defects (CDs) produced by heavy ion
irradiation is one of the most intriguing questions in the current
study of the vortex state in high temperature superconductors.  For
the most anisotropic Bi$_2$Sr$_2$CaCu$_2$O$_{8-\delta}$ (Bi-2212)
superconductor without strong disorder, neutron scattering and
Josephson plasma resonance (JPR) data provide evidence in favor of a
pancake liquid with very weak correlations of pancakes in different
layers \cite{neut,jpr,k}.  When CDs are introduced into these crystals
a large decrease in the reversible magnetization is observed
\cite{drost-kees}, indicating that pancakes are predominantly situated
on the CDs, even in the liquid state.  Recent studies of JPR
\cite{kos,tam} and $c$-axis transport \cite{mor,bvm} in irradiated
Bi-2212 crystals reveal enhancement of $c$-axis correlations in some
interval of out-of-plane magnetic fields, $B_{\bot}$, below the
matching field, $B_{\Phi}$, at temperatures $T \sim 70$~K. 
In other words, pancakes at these fields and temperatures appear
to form aligned segments of vortex lines inside CDs, while outside of
this region they are better described as a liquid of weakly
$c$-axis-correlated pancakes, as in pristine crystals.  Both $c$-axis
transport and JPR involve the flow of Josephson currents and are thus
sensitive to the misalignment of pancake vortices between adjacent
planes.  CDs aligned with the $c$-axis will promote interplane pancake
alignment in a region of temperature where pancakes remain largely
localized on CDs.  Even in the liquid state, where pancakes are
mobile, filling of the available CD sites should lead to enhanced
$c$-axis correlation from statistical considerations alone.  In
previous $c$-axis transport measurements \cite{mor}, a dip in the
magnetoresistance, $\rho(B_{\bot})$, of a Bi-2212 crystal was observed
at a magnetic field corresponding to a filling factor of CDs, $f 
=B_{\bot}/B_{\Phi}\approx 1/3$.

So far there has been no technique available for providing 
quantitative information on the degree of c-axis correlation or on the 
length of pancake line segments.  In addition, it is unclear whether 
vortex interactions play a significant role in enhancing correlation 
in the liquid state.  Previously, a method was proposed, but not yet 
realized, to extract the pancake density correlation function using 
data for JPR frequency as a function of $B_{\parallel}$ at fixed 
$B_{\bot}$ \cite{kbm}.  In this Letter we determine the average length of 
vortex line segments as a function of $B_{\bot}$, i.e.  of the filling 
factor $f$, from transport measurements on single crystals of Bi-2212 
with and without CDs.  For this we have developed a method for 
extracting the phase difference correlation function along the layer 
from measurements of the $c$-axis conductivity, $\sigma_c$, as a 
function of the magnetic field component parallel to the layers, 
$B_{\parallel}$, at fixed $B_{\bot}$.  The component $B_{\bot}$ 
establishes the vortex state to be studied, while the component 
$B_{\parallel}$ serves as a probe of this state, as described below.  
Knowing the phase difference correlation length, we estimate the 
pancake density correlation length along the $c$ axis.

In Josephson coupled superconductors in the presence of a $c$-axis
current, voltage is induced by slips of the phase difference between
layers, $\varphi_{n,n+1}({\bf r},t)$, as described by the Josephson
relation $V_{n,n+1}=(\hbar/2e)\dot{\varphi}_{n,n+1}$.  Here $n$
labels
layers, ${\bf r}=x,y$ are coordinates in the $ab$ plane, and $t$ denotes
the time.  The $c$-axis conductivity in the vortex liquid state is
determined by the Kubo formula \cite{k}
\begin{eqnarray}
\sigma_c(B_{\bot},B_{\parallel})=(s{\cal J}_0^2/2T)\int_0^{\infty}dt
\int d{\bf r}G({\bf r},t),
\label{sig} \\
G({\bf r},t)=2\langle\sin\varphi_{n,n+1}(0,0)
\sin\varphi_{n,n+1}({\bf r},t)\rangle
\nonumber \\
\approx\langle\cos[\varphi_{n,n+1}({\bf r},t)-\varphi_{n,n+1}(0,0)]\rangle,
\label{1}
\end{eqnarray}
where ${\cal J}_0$ is the Josephson critical current and $s$ is the
interlayer distance.  Time variations of
the phase difference are caused by mobile
pancakes \cite{k1} induced by $B_{\bot}$ and by mobile Josephson vortices
induced by the parallel component $B_{\parallel}$.
In the lowest order in Josephson coupling we split
$[\varphi_{n,n+1}(0,0)-\varphi_{n,n+1}({\bf r},t)]$ into the
contribution induced by pancakes and that caused by the unscreened
parallel
component $B_{\parallel}$. Assuming that $B_{\parallel}$ is along
the $x$ axis, we obtain a simple expression for the contribution of
the parallel component to the phase difference:
\begin{eqnarray}
&&\varphi_{n,n+1}(0,0)-\varphi_{n,n+1}({\bf r},t)\approx \nonumber\\
&&[\varphi_{n,n+1}(0,0)-\varphi_{n,n+1}({\bf
r},t)]_{B_{\parallel}=0}-
2\pi sB_{\parallel}y/\Phi_0.
\label{me}
\end{eqnarray}
Inserting this expression into Eqs.~(\ref{sig}) and (\ref{1}) we
obtain
\begin{equation}
\sigma_c(B_{\bot},B_{\parallel})=(\pi s{\cal J}_0^2/T)\int
drr\tilde{G}(r,B_{\bot})
J_0(\alpha B_{\parallel}r),
\label{me1}
\end{equation}
where $J_0(x)$ is the Bessel function, $\alpha=2\pi s/\Phi_0$ and
\begin{equation}
\tilde{G}({\bf r},B_{\bot})=\int_0^{\infty} dtG({\bf r},t,B_{\bot}).
\end{equation}
The function $G({\bf r},t)$ describes the dynamics of the phase
difference caused by mobile pancakes.  If
$g(B_{\bot},B_{\parallel})=\sigma_c(B_{\bot},B_{\parallel})/\sigma_c(B_{\bot},0)
$ is
known in the vortex liquid, the correlation function
$G(r)=\tilde{G}(r){\cal J}_0^2\Phi_0^2/4\pi sT\sigma_c(B_{\bot},0)$ may be
found using the inverse Fourier-Bessel transform:
\begin{equation}
G(r,B_{\bot})=\int dB_{\parallel}
B_{\parallel}g(B_{\parallel},B_{\bot})J_0(\alpha
B_{\parallel}r).
\label{tr}
\end{equation}
The correlation lengths $R$ and $R_1$ of this function are defined by
the relations:
\begin{equation}
R^2=\int_0^{\infty}dr\frac{rG(r)}{G(0)}, \ \ \
R_1^2=\int_0^{\infty}dr\frac{r^3G(r)}{R^2G(0)}.
\label{22}
\end{equation}
Note  that $R_1$
is related to the coefficient of the $B_{\parallel}^2$ term in the
expansion of $g(B_{\bot},B_{\parallel})$ in $B_{\parallel}^2$:
\begin{equation}
g(B_{\bot},B_{\parallel}) \approx 1-[\pi
sR_1(B_{\bot})/\Phi_0]^2B_{\parallel}^2.
\label{rel}
\end{equation}
Thus $R_1(B_{\bot})$ can be obtained independently from data for
$\sigma_c(B_{\bot},B_{\parallel})$ at small $B_{\parallel}$.

Such a procedure to obtain the function $G(r,B_{\bot})$ is valid if
the vortex state depends weakly on the Josephson coupling and hence on
the probe field $B_{\parallel}$ (which affects Josephson coupling).
Then this method is nondestructive.  Let us check first under what
conditions we can neglect the effect of Josephson coupling on the
equilibrium vortex state.  The energy of Josephson coupling in the
correlated area $\pi R^2$ is $\pi E_0R^2/\lambda_J^2$, which should be
much smaller than the temperature $T$ to be treated as a perturbation
\cite{kbm}.  Here $E_0(T)=\Phi_0^2s/16\pi^3\lambda_{ab}^2(T)$.  For
Bi-2212, with $\gamma\approx 300$ and $\lambda_{ab}(0)\approx 2000$
\AA, the Josephson coupling in the correlated area is $\approx 0.2T$
at the maximum value of $R$ found below and at $T> 60$ K. Thus the
effect of Josephson coupling, and hence $B_{\parallel}$, on the
equilibrium vortex state may be neglected.  For dynamical parameters,
such as $\sigma_c$, higher order terms in Josephson coupling
describing dynamic screening of $B_{\parallel}$ omitted in Eq.~(3),
may be important.  This will be discussed below.

We anticipate that in the pancake vortex liquid state in pristine
crystals the characteristic lengths $R$ and $R_1$ of the correlation
function $G(r)$ are of order of the intervortex distance,
$a=(\Phi_0/B_{\bot})^{1/2}$, because each pancake here is mobile and
induces phase slippage as described in Ref.~\onlinecite{k1}.  In
irradiated crystals with CDs we anticipate much bigger $R$ and $R_1$
if pancakes form long segments of lines inside CDs.  Then only ends of
segments contribute significantly to the phase difference
$\varphi_{n,n+1}({\bf r},t)$ and lead to suppression of Josephson
coupling and $\sigma_c$, while the effect of pancakes in neighboring
layers inside the segments is much smaller.

For our experiments, high quality Bi-2212 crystals $(T_{c}\simeq {\rm
85\:K})$ of about $1\times 1.5\times 0.02$ mm$^3$ were used.  The
irradiation by 1.2 GeV ${\rm U^{238}}$-ions was performed on the ATLAS
accelerator (ANL).  According to TRIM calculations these ions produce
in Bi-2212 continuous amorphous tracks with diameter 4-8~nm and length
25-30 $\mu$m.  Below, we present the results for the samples
irradiated with a density of CDs corresponding to the matching field
$B_{\Phi }=2$~T and for a reference pristine sample.

Our measurements of $c$-axis conductivity were carried out in a
cryostat with {\em two} superconducting magnets providing magnetic
fields in orthogonal directions.  The magnets are controlled
independently and provide fields up to 8 T in one direction and up to
1.5 T in the other direction.  Samples can be oriented along the axis
of either magnet.  Misalignment of the crystal $c$ axis with respect
to the perpendicular component $B_{\bot}$ was detected by the
asymmetry of $\rho_c(B_{\parallel})$ with respect to the sign of
$B_{\parallel}$.  We adjusted the direction of the field components,
providing asymmetry below 5\% at $B_{\parallel}>2$~T and below 10\% at
lower fields.  The normalized conductivity $g(B_{\parallel})$ was 
calculated using the average resistivity 
$\overline{\rho}_c(B_{\parallel}) =[\rho_c(B_{\parallel}) 
+\rho_c(-B_{\parallel}) ]/2$.  Two silver contact pads were deposited 
on both sides of the sample using a mechanical shadow mask.  The mask 
provided a clean surface rim $\sim 0.1$~mm from the sample edges and 
${\rm 25\:\mu m}$ separation between current and potential pads.  The 
area of the contact  was $\approx 0.75 \:{\rm mm^2}$ for current 
and $\approx 0.05 \: {\rm mm^2}$ for potential terminals.  The 
resistance of the current pads at room temperature was $\approx 
2$~$\Omega$.  A current of 1~mA, driven through the sample, provides 
an ohmic I-V regime.  For very anisotropic Bi-2212, in our range 
of the magnetic field and temperatures, in-plane conductivity, 
$\sigma_{ab}\sim 10^3 \: \sigma_c$ \cite{richard}, providing nearly 
equipotential current distribution in the $ab$-plane, at least in the 
central area of the sample.  Thus the contribution of  
$\sigma_{ab}$ to the anisotropic conductivity is weak and can be 
neglected, in the limit of small pad separation.  Thus a standard 
4-probe method can be used for $\rho_c$ measurement instead of the 
complicated multiterminal Montgomery analysis.  The temperature was 
stabilized with an accuracy $\pm 50$~mK. The resistivity 
$\rho_{c}(B_{\parallel})$  was measured in 
the $B_{\perp}$ interval where $g$ at maximum $B_{\parallel}$ drops 
with $B_{\parallel}$ at least to the value 0.2.

In Fig.~1 we present $\rho_c$ as a function of $B_{\parallel}^2$ at
different $B_{\perp}$ for a) irradiated and b) pristine crystals.  For
irradiated crystals at low $B_{\parallel}$ the resistance increases
quadratically with $B_{\parallel}$ as prescribed in Eq.~(8).  In
contrast, for the pristine crystal we observed that $\rho_c$ increases
quadratically at high $B_{\parallel}$, but exhibits a minimum at low
fields that is not described by Eq.~(8).  This low field behavior is
caused by dynamic screening of $B_{\parallel}$ that results in
additional dissipation at low fields, due to the combined effects of
motion of Josephson vortices induced by $B_{\parallel}$ and of pancake
vortices.  These effects are not important at high fields.  They do
not appear in the irradiated crystal due to pinning of the Josephson
vortices.

%%%%%%%%%%%%%%%%%%%%%%%%%%%%%%%%%%%%%%%%%%%%%%%%%%%%%%%%%%%%%%%%%%%%%%%%%%%%
																		   %
	\begin{figure}[h]													   %
	\vspace{-0.5cm}														   %
	\rightline{	\epsfxsize = 9cm \epsffile{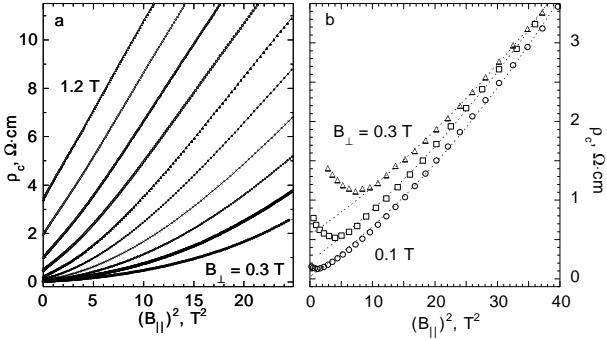}}					   %
	\vspace{-1cm}														   %
	\caption {															   %
	Dependence of resistivity $\rho_c$ versus $B_{\parallel}^{2}$		   %
	for	perpendicular components $B_{\bot}$, increasing	sequentially	   %
	 with the step 0.1~T  in irradiated	(a), and pristine (b) crystals.}   %
	\end{figure}														   %
																		   %
%%%%%%%%%%%%%%%%%%%%%%%%%%%%%%%%%%%%%%%%%%%%%%%%%%%%%%%%%%%%%%%%%%%%%%%%%%%%

In Fig.~2 we show the dependence $g=\sigma_{c}(b)/\sigma_c(0)$ on
$b=B_{\parallel}/(B_sB_{\bot})^{1/2}$ for irradiated and pristine
samples at different $B_{\bot}$.
Here $B_s=\Phi_0/2\pi s^2$.  For the pristine crystal the values
$\sigma_c(B_{\bot})$ at $B_{\parallel}=0$
were determined by extrapolation of $\rho_c(B_{\parallel})$ from the
high field quadratic dependence to zero  as shown in Fig.~1b by the
dashed lines. Note that for the pristine crystal all three
curves

%%%%%%%%%%%%%%%%%%%%%%%%%%%%%%%%%%%%%%%%%%%%%%%%%%%%%%%%%%%
														  %
	\begin{figure}[h]									  %
	\rightline{	\epsfxsize = 8.5cm \epsffile{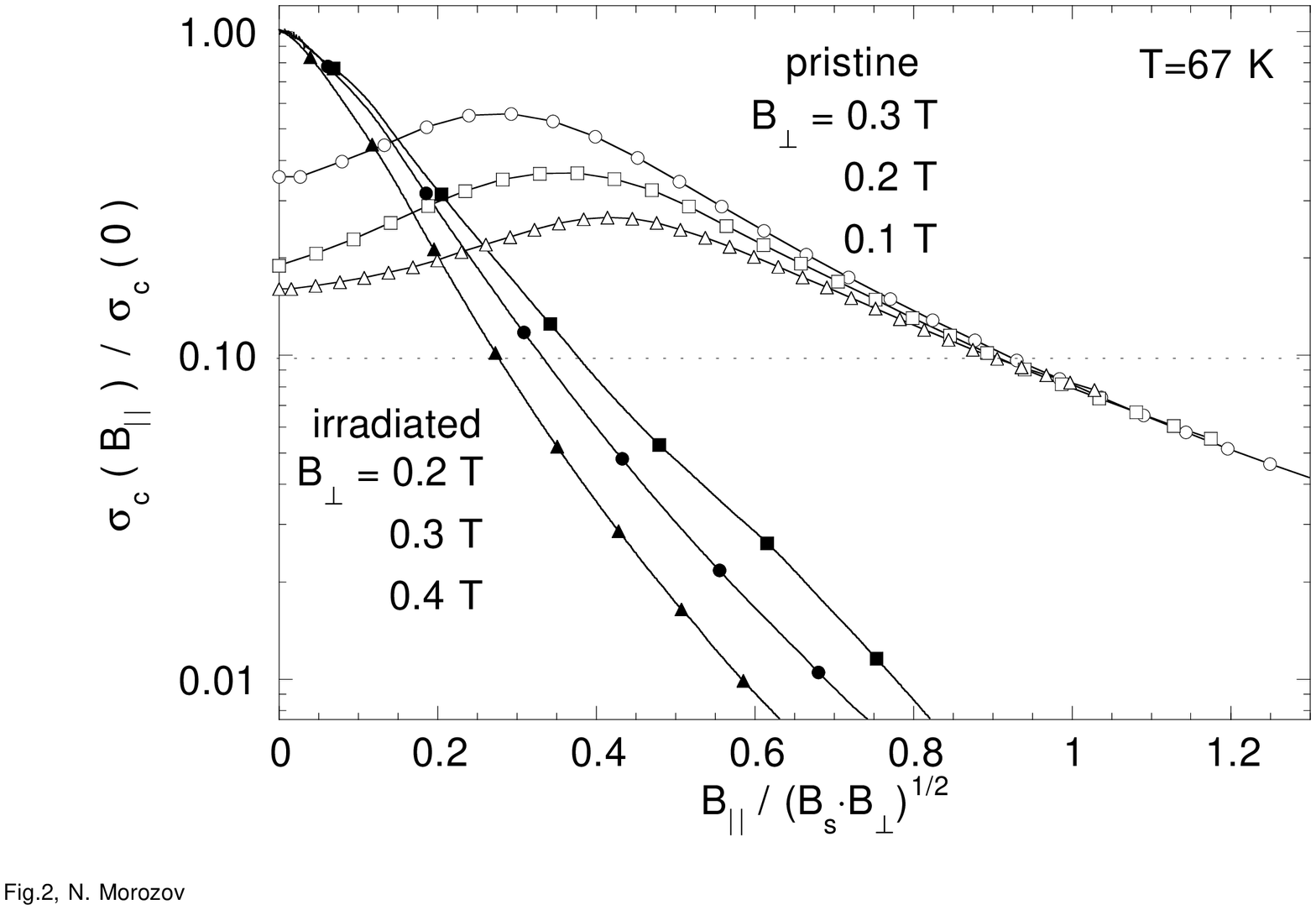}}	  %
	\caption{Dependence	$g(B_{\parallel})=\sigma_{c}	  %
	[B_{\parallel}/(B_sB_{\bot})^{1/2}]/\sigma_c(0)$	  %
	in  irradiated and   pristine		          
	crystals.}											  %
	\end{figure}										  %
														  %
%%%%%%%%%%%%%%%%%%%%%%%%%%%%%%%%%%%%%%%%%%%%%%%%%%%%%%%%%%%

\noindent coincide at high fields, demonstrating scaling of the 
correlation length $R$ with $a$.  Such a scaling is not observed for 
the irradiated crystal at $B_{\bot}\leq 1.2$ T, because the average 
distance between CDs, $a_{\Phi}=(\Phi_0/B_{\Phi})^{1/2}$, gives 
another length scale in addition to $a$.  More importantly, we see 
that $g$ drops with $B_{\parallel}$ much faster for the irradiated 
sample.  The correlation length $R(B_{\bot})$ is related via $R\approx 
\Phi_0/\tilde{B}_{\parallel}s$ to the magnetic field 
$\tilde{B}_{\parallel}$ which characterizes the scale of the drop of 
$\sigma_c$ with $B_{\parallel}$.  Here $\tilde{B}_{\parallel}$ is the 
magnetic field at which flux in the area $Rs$ is $\approx \Phi_0$.  If 
we define the characteristic field ${\tilde B}_{\parallel}$, as given 
by: $g({\tilde B}_{\parallel})=0.1$, then from Fig.2 at 
$B_{\perp}=0.2$~T, we estimate for the pristine crystal $R/a \approx 
2.5$, while for irradiated one $R/a\approx 8$.  As one can see from 
Fig.~2, at the high fields $B_{\parallel}$ used in our measurements, 
the dependence $g(B_{\parallel})$ is close to exponential, and we use 
this to extrapolate $g(B_{\parallel})$ to higher fields.  Then we 
determine $G(r)$ in the interval $r<r_m\approx 0.5$ $\mu$m by use of 
Eq.~(\ref{tr}).  Accuracy of our experimental data and of the inverse 
Fourier-Bessel transformation is not sufficient to obtain $G(r)$ at 
$r>r_m$.

%%%%%%%%%%%%%%%%%%%%%%%%%%%%%%%%%%%%%%%%%%%%%%%%%%%%%%%%%%%%%%%%%%%%%
																	%
	\begin{figure}[h]												%
	\vspace{-0.5cm}													%
	\hspace{-0.5cm}													%
	\epsfxsize = 9.5cm \epsffile{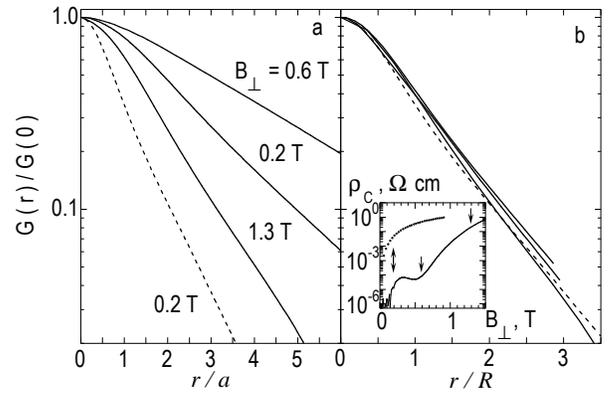}						%
	\caption{The functions $G(r/a)$	(left),	and	$G(r/R)$ (right)	%
	as extracted from $g(B_{\parallel})$							%
	for	irradiated (solid) and for pristine	(dashed) samples.   	%
	Inset: $\rho_c(B_{c})$ at										%
	$B_{\parallel}=0$ from Ref.~7.	Arrows							%
	indicate $B_{\bot}$	values for the presented $G(r)$	curves.}	%
	\end{figure}													%
																	%
%%%%%%%%%%%%%%%%%%%%%%%%%%%%%%%%%%%%%%%%%%%%%%%%%%%%%%%%%%%%%%%%%%%%%

In Fig.~3a we show the functions $G(r/a)$  
and in Fig.~3b the same data 
versus $x=r/ R$, where the scaling length $R(B_{\bot}, B_{\Phi})$ is 
defined by Eq.~\ref{22}.  For the pristine crystal at $B_{\bot}=0.1$~T 
and 0.2~T we obtain $R(B_{\bot})\approx a$.  The curve for 
$B_{\bot}=0.2$~T is plotted by dashed line and it practically 
coincides with that at $B_{\bot}=0.1$~T. For the irradiated sample we 
show the curves corresponding to 
regions below,  above, and at the position of the dip in the 
magnetoresistance curve, $\rho_c(B_{\bot},0)$ from Ref.~\cite{mor} 
as shown in the inset. It 
is evident in  Fig.~3a that the rate of decay of correlations with 
$r$ is not a monotonic function of $B_{\bot}$ and is a minimum at 
a value corresponding to the dip in $\rho_c(B_{\bot})$.
Notably  all these curves quite accurately  merge in Fig~3b, 
providing {\em a universal function} $G(x)$ for both pristine and 
irradiated samples with a single scaling length $R$, although the 
irradiated sample is characterized generally by two lengths $a$ and 
$a_{\Phi}$. This means that $G(x)$ is determined mainly by the static 
effect of vortices on Josephson coupling \cite{k} though it is 
extracted from the dynamic quantity $\sigma_c$.  In comparison with 
pristine crystals CDs simply diminish the effective concentration of 
pancakes acting on Josephson coupling by the factor $a^2/R^2$.

The correlation length R as a
function of $B_{\bot}$ is shown in Fig.~4.
It exhibits a distinct maximum,
$R/a\approx 4$, at $f\approx 0.35$, again coinciding with the position
of the dip in $\rho_c(B_{\bot})$. At $f=0.35$ the ratio of
$R(B_{\bot})$ for irradiated and pristine crystals is about 4 times.

%%%%%%%%%%%%%%%%%%%%%%%%%%%%%%%%%%%%%%%%%%%%%%%%%%%%%%%%%%%%%%%%%%%%%%%%%%%%%%%%%%%%%%
																					 %
	\begin{figure}[h]																 %
	\rightline{	\epsfxsize = 7.7cm \epsffile{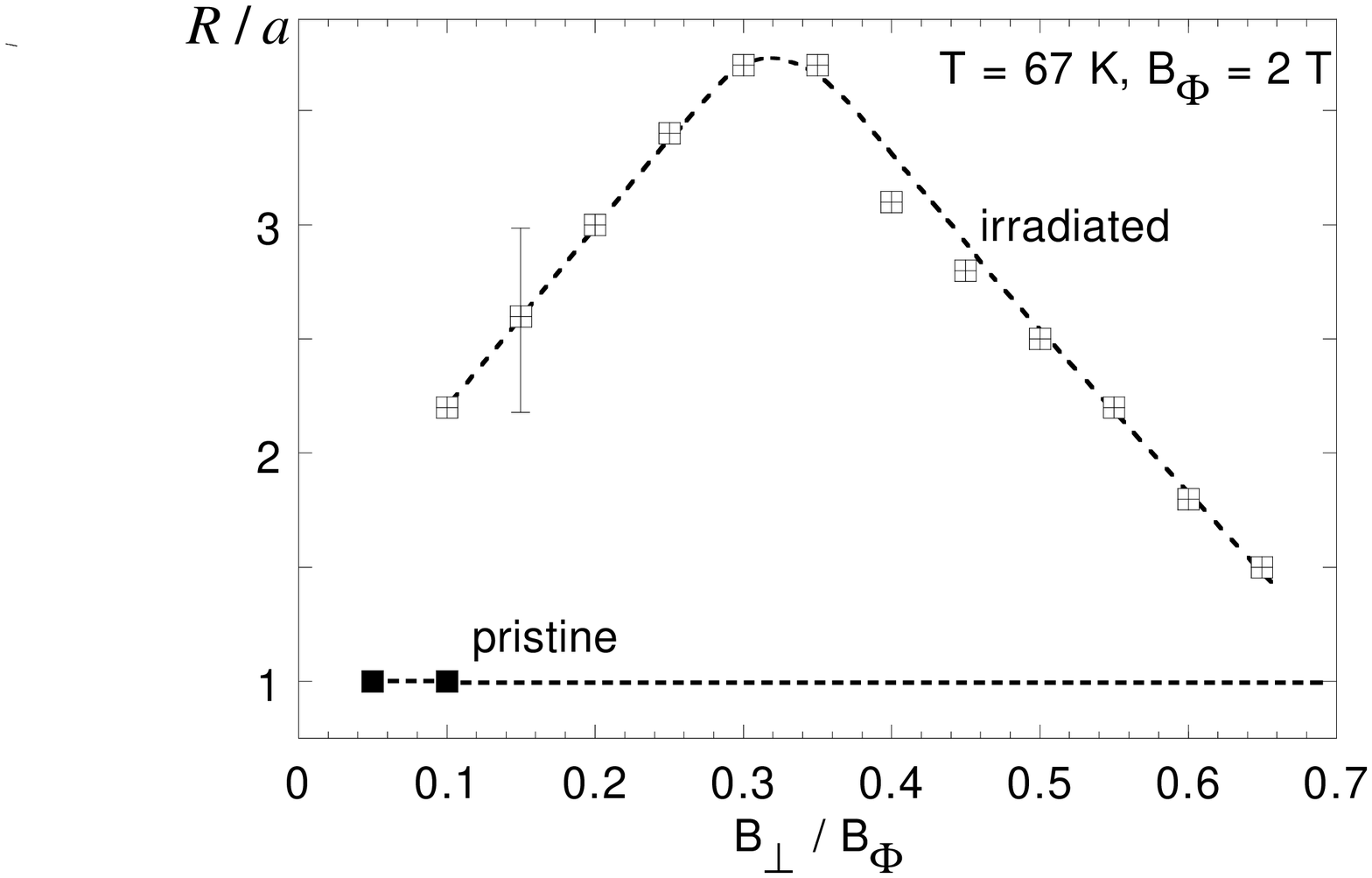}}								 %
	\caption{Dependence	of the correlation length $R$								 %
	as a function of filling factor, $f=B_{\bot}/B_{\Phi}$,	of columnar				 %
	defects	 for irradiated	and	 pristine samples. Dashed lines	are	a guide.}  	   %
	\end{figure}																	 %
																					 %
%%%%%%%%%%%%%%%%%%%%%%%%%%%%%%%%%%%%%%%%%%%%%%%%%%%%%%%%%%%%%%%%%%%%%%%%%%%%%%%%%%%%%%

The length $R_1$ obtained from the function $G(r)$ using Eq.~(7) is
$\approx 2R$ at all values $B_{\bot}$ studied.  The same length
determined directly from the function $g(B_{\parallel})$ at small
$B_{\parallel}$ using Eq.~(\ref{rel}) is $\approx 3R$. Thus we
estimate the accuracy of extracting $G(r)$ as $\approx 30$~\%.

In order to determine the $c$-axis correlations of pancakes
and to explain scaling, we note that at small $f$ pancakes are positioned
mainly inside CDs and hence the drop in the phase difference
correlations is caused by interruptions in pancake arrangement inside
CDs.   Namely, we suppose that the phase difference
$\varphi_{n,n+1}({\bf r})$ is induced when a columnar defect in the
layer $n$ is occupied by a pancake, but the site inside the same CD in
the layer $n+1$ is empty (and vise versa). At larger $f$ the unpinned pancakes also contribute to the
 decay of correlations. Scaling for both pristine and 
irradiated crystals means that ends of vortex segments inside CDs act 
on the phase difference correlations as unpinned pancakes. The net 
concentration of interruptions and unpinned pancakes is $1/R^2$, i.e. 
the average length of vortex segments is $L/s\approx R^2/a^2$. 
In the most aligned vortex liquid, at $f\approx
0.35$, we estimate $L/s\approx 15 $.  This is much larger than 
$L_0/s\approx 1/(1-f)\approx 1.5$ in the model of
noninteracting pancakes positioned randomly on CDs.  Thus we conclude
that interaction of pancakes is important for enhanced alignment
inside CDs.

In the hierarchy of interactions in the presence of CDs, both the
pinning energy per pancake and the intralayer repulsion energy of
pancakes are characterized by the same energy scale $E_0$.  The scale
of magnetic pair attraction of pancakes in different layers is smaller
by the factor $s/\lambda_{ab}$.
The random distribution of CDs is important.  Repulsion between vortices in the same layer
leads to a significant increase of pancake energy inside CDs situated
near those already occupied by pancakes \cite{buz}.  As $B_{\bot}$
increases, some CDs become more favorable for filling by pancakes,
while  others remain unoccupied.  Another important point is
that favorable configurations are similar in all layers due to the
geometry of CDs.  Thus repulsion of pancakes inside randomly
positioned CDs in the same layer leads to enhancement of $c$-axis
correlations.  Another mechanism for pancake alignment inside CDs is
magnetic attraction of pancakes in adjacent layers. 

In conclusion, we extract the universal phase difference correlation function
using measurements of $c$-axis resistivity as a function of the
parallel component of the magnetic field at fixed perpendicular
component.  We estimated the correlation length of the
pancake density correlation function along the $c$ axis as a function
of the filling factor $f$ of columnar defects.  It first increases
 with $f$, reaches a maximum at $f\approx 0.35$, and then
drops as pancakes start to occupy positions outside of CDs.
 We argue that enhancement of
the alignment of pancakes inside columnar defects is caused by
interaction of pancakes confined inside columnar defects.

Useful discussions with V.M.~Vinokur are greatly appreciated.  We
thank P.~Kes and T.-W.~Li for providing Bi-2212 single crystals and
R.~Olsson for technical assistance.  This work was supported by the
U.S. DOE.

\end{document}